\documentclass[12pt]{JHEP3}
\usepackage{epsfig}

\newcommand{\ra}{\rightarrow}

\newcommand{\bc}{\begin{center}}
\newcommand{\ec}{\end{center}}
\newcommand{\ba}{\begin{array}}
\newcommand{\ea}{\end{array}}
\newcommand{\beq}{\begin{equation}}
\newcommand{\eeq}{\end{equation}}
\newcommand{\bea}{\begin{eqnarray}}
\newcommand{\eea}{\end{eqnarray}}
\newcommand{\bmx}{\begin{pmatrix}}
\newcommand{\emx}{\end{pmatrix}}
\newcommand{\nn}{\nonumber}

\newcommand{\dl}{\delta}

\newcommand{\n}{\nu}

\newcommand{\te}{\theta}

\newcommand{\del}{\partial}
\newcommand{\half}{\frac{1}{2}}
\newcommand{\tr}{{\rm tr}}
\newcommand{\tbar}{{\overline t}}

\newcommand{\sref}[1]{\S\,\ref{#1}}
\newcommand{\eref}[1]{Eq.~(\ref{#1})}

\newcommand{\cZ}{{\cal Z}}
\newcommand{\zbar}{{\bar z}}
\newcommand{\cT}{{\cal T}}
\newcommand{\tA}{{\tilde A}}
\newcommand{\blangle}{\bigg\langle}
\newcommand{\brangle}{\bigg\rangle}
\def\IB{\relax{\rm I\kern-.18em B}}
\def\IC{{\relax\hbox{\kern.3em{\cmss I}$\kern-.4em{\rm C}$}}}
\def\ID{\relax{\rm I\kern-.18em D}}
\def\IE{\relax{\rm I\kern-.18em E}}
\def\IF{\relax{\rm I\kern-.18em F}}
\def\II{\relax{\rm I\kern-.18em I}}
\def\Id{\relax{1\kern-.32em 1}}
\def\IG{\relax\hbox{$\inbar\kern-.3em{\rm G}$}}
\def\IR{\relax{\rm I\kern-.18em R}}

\title{$c=1$ Matrix Models: Equivalences and Open-Closed String Duality}
\author{Anindya Mukherjee\,\footnote{Email: anindya\_m@theory.tifr.res.in}\,
and Sunil Mukhi\,\footnote{Email: mukhi@tifr.res.in}\\
\it Tata Institute of Fundamental Research,\\
\it Homi Bhabha Rd, Mumbai 400 005, India}
\abstract{We give an explicit demonstration of the equivalence 
between the Normal Matrix Model (NMM) of $c=1$ string theory at
selfdual radius and the Kontsevich-Penner (KP) model for the same
string theory. We relate macroscopic loop expectation values in the
NMM to condensates of the closed string tachyon, and discuss the
implications for open-closed duality. As in $c<1$, the Kontsevich-Miwa
transform between the parameters of the two theories appears to encode
open-closed string duality, though our results also exhibit some
interesting differences with the $c<1$ case. We also briefly comment
on two different ways in which the Kontsevich model originates.}

\preprint{hep-th/0505180\\ TIFR/TH/05-18}

\keywords{String theory, Random matrices}

\begin{document}

\section{Introduction}
\label{intro}

In the last few years, enormous progress has been made in
understanding noncritical string theory. One line of development
started with the work of
Refs.\cite{McGreevy:2003kb,Martinec:2003ka,Klebanov:2003km}, in the
context of D-branes of Liouville theory.  These and subsequent works
were inspired by the beautiful CFT computations that gave convincing
evidence for the consistency of these
branes\cite{Zamolodchikov:2001ah,Fateev:2000ik,Teschner:2000md}, as
well as Sen's picture of the decay of unstable D-branes via tachyon
condensation\cite{Sen:2004nf}. Another independent line of
development that has proved important was the attempt to formulate new
matrix models to describe noncritical string theories and their
deformations, including black hole
deformations\cite{Kazakov:2000pm,Alexandrov:2001cm,%
Alexandrov:2002fh,Alexandrov:2003qk}.
 
Some of the important new results are related to nonperturbatively
stable type 0 fermionic strings\cite{Takayanagi:2003sm,
Douglas:2003up}, but even in the bosonic context, many old and new
puzzles concerning matrix models as well as Liouville theory have been
resolved.  For $c<1$ matter coupled to Liouville theory, a beautiful
picture emerged of a Riemann surface governing the semiclassical
dynamics of the model. Both ZZ and FZZT branes were identified as
properties of this surface: the former are located at singularities
while the latter arise as line integrals. This picture was obtained in
Ref.\cite{Seiberg:2003nm} within the continuum Liouville approach and
subsequently re-derived in the matrix model formalism in
Ref.\cite{Kazakov:2004du} using earlier results of
Ref.\cite{Daul:1993bg}. However, later it was
realised\cite{Maldacena:2004sn} that the exact, as opposed to
semiclassical, picture is considerably simpler: the Riemann surface
disappears as a result of Stokes' phenomenon and is replaced by a
single sheet. In the exact (quantum) case, correlation functions of
macroscopic loop operators go from multiple-valued functions to the
Baker-Akhiezer functions of the KP hierarchy, which are analytic
functions of the boundary cosmological constant.  Thus, for these
models (and also their type 0 extensions) a rather complete picture
now exists.

Another remarkable development in this context is an explicit proposal
to understand open-closed string duality starting from open string
field theory. This was presented in Ref.\cite{Gaiotto:2003yb} and
implemented there for the $(2,q)$ series of minimal models coupled to
gravity (which can be thought of as perturbations of the ``topological
point'' or $(2,1)$ minimal model). The basic idea of
Ref.\cite{Gaiotto:2003yb} was to evaluate open string field theory on
a collection of $N$ FZZT branes in the $(2,1)$ closed string
background. This leads to the Kontsevich matrix
model\cite{Kontsevich:1992ti}, which depends on a constant matrix $A$
whose eigenvalues are the $N$ independent boundary cosmological
constants for this collection of branes. Now the Kontsevich model
computes the correlators of {\it closed-string} observables in the
same $(2,1)$ background. So this relationship was interpreted as
open-closed duality, following earlier ideas of Sen\cite{Sen:2003iv}.

A different way of understanding what appears to be the same
open-closed duality emerged in Ref.\cite{Maldacena:2004sn} for the
$(2,1)$ case. Extending some older observations in
Ref.\cite{Moore:1991ir}, the authors showed that if one inserts
macroscopic loop operators $\det(x_i-\Phi)$, representing FZZT branes
(each with its own boundary cosmological constant $x_i$) in the
Gaussian matrix model, and takes a double-scaling limit, one obtains
the Kontsevich matrix model. The constant matrix $A$ in this model
again arises as the boundary cosmological constants of the FZZT
branes\footnote{This has been generalised\cite{Hashimoto:2005bf} by
starting with macroscopic loops in the double-scaled 2-matrix models
that describe $(p,1)$ minimal model strings. After double-scaling, one
obtains the generalised Kontsevich models of
Refs.\cite{Adler:1992tj,Kharchev:1991cy}.}.

The situation is more complicated and less well-understood for $c=1$
matter coupled to Liouville theory, namely the $c=1$ string. The
results of FZZT were derived for generic Liouville central charge
$c_L$, but become singular as $c_L\to 25$, the limit that should give
the $c=1$ string. Attempts to understand FZZT branes at $c=1$
(Refs.\cite{Ghoshal:2004mw,Alexandrov:2005gf}) rely on this limit from
the $c<1$ case which brings in divergences and can therefore be
problematic. In particular, there is as yet no definite computation
exhibiting open-closed duality at $c=1$ starting from open string
field theory in the $c=1$ Liouville background. One should expect such
a computation to give rise to the $c=1$ analogue of the Kontsevich matrix
model, namely the Kontsevich-Penner model\footnote{This model is valid only at the
selfdual radius $R=1$.} of Ref.\cite{Imbimbo:1995yv}.

In the present work we take a different approach to understand
D-branes and open-closed duality in the $c=1$ string, more closely
tied to the approach of
Refs.\cite{Maldacena:2004sn,Hashimoto:2005bf}. The obvious point of
departure at $c=1$ would be to consider macroscopic loops in the
Matrix Quantum Mechanics (MQM) and take a double-scaling
limit. Indeed, FZZT branes at $c=1$ have been investigated from this
point of view, for example in
Refs.\cite{Maldacena:2005hi,Gaiotto:2005gd}.  However, we will take an
alternative route that makes use of the existence of the Normal Matrix
Model (NMM)\cite{Alexandrov:2003qk} for $c=1$ string theory (in
principle, at arbitrary radius $R$). This model is dual in a certain
precise sense to the more familiar MQM, namely, the grand canonical
partition function of MQM is the partition function of NMM in the
large-$N$ limit. Geometrically, the two theories correspond to
different real sections of a single complex curve. More details about
the interrelationship between MQM and NMM can be found in
Ref.\cite{Alexandrov:2003qk}.

One good reason to start from the NMM is that it is a simpler model
than MQM and does not require a double-scaling limit. Also, it has
been a longstanding question whether the KP model and NMM are
equivalent, given their structural similarities, and if so, what is
the precise map between them. It is tempting to believe that
open-closed duality underlies their mutual relationship. Indeed, the
NMM does not have a parameter suggestive of a set of boundary
cosmological constants, while the KP model has a Kontsevich-type
constant matrix $A$. So another natural question is whether the
eigenvalues of $A$ are boundary cosmological constants for a set of
FZZT branes/macroscopic loop operators of NMM.

In what follows we examine these questions and obtain the following
results. First of all we find a precise map from the NMM (with
arbitrary tachyon perturbations) to the KP model, thereby
demonstrating their equivalence. While the former model depends on a
non-Hermitian matrix $Z$ constrained to obey $[Z,Z^\dag]=1$, the
latter is defined in terms of a positive definite Hermitian matrix
$M$. We find that the eigenvalues $z_i$ and $m_i$ are related by
$m_i=z_i\zbar_i$. The role of the large-$N$ limit in the two models is
slightly different: in the KP model not only the random matrix but
also the number of parameters (closed string couplings) is reduced at
finite $N$. On the contrary, in the NMM the number of parameters is
always infinite for any $N$, but one is required to take $N\to\infty$
to obtain the right theory (this was called ``Model I'' in
Ref.\cite{Alexandrov:2003qk}). The two models are therefore equivalent
only on a subspace of the parameter space at finite $N$, with the
limit $N\to\infty$ being required to obtain full equivalence. This is
an important point to which we will return.

Next in \sref{loopops} we consider macroscopic loop operators of the
form $\det(\xi-Z)$ in the NMM, and show that these operators when
inserted into the NMM, {\it decrease} the value of the closed-string
tachyon couplings in a precise way dictated by the Kontsevich-Miwa
transform. On the contrary, operators of the form $1/\det(\xi-Z)$ play
the role of increasing, or turning on, the closed-string tachyon
couplings. In particular, insertion of these {\it inverse determinant}
operators in the (partially unperturbed) NMM leads to the
Kontsevich-Penner model. (By partially unperturbed, we mean the
couplings of the positive-momentum tachyons are switched off, while
those of the negative-momentum tachyons are turned on at arbitrary
values.) Calculationally, this result is a corollary of our derivation
of the KP model from the perturbed NMM in \sref{Equivalence}.

These results bear a rather strong analogy to the emergence of the
Kontsevich model from the insertion of determinant operators at
$c<1$\cite{Maldacena:2004sn}. In both cases, the parameters of
macroscopic loop operators turn into eigenvalues of a Kontsevich
matrix. Recall that in Ref.\cite{Maldacena:2004sn}, one inserts $n$
determinant operators into the $N\times N$ Gaussian matrix model and
then integrates out the Gaussian matrix. Taking $N\to\infty$ (as a
double-scaling limit) we are then left with the Kontsevich model of
rank $n$. In the $c=1$ case, we insert $n$ inverse determinant
operators in the NMM. As we will see, $N-n$ of the normal matrix
eigenvalues then decouple, and we are left with a Kontsevich-Penner
model of rank $n$ (here one does not have to take $N\to\infty$). We
see that the two cases are rather closely analogous.

The main difference between our case at $c=1$ and the $c<1$ case of
Ref.\cite{Maldacena:2004sn} is that we work with inverse determinant
rather than determinant operators. However at infinite $n$ we can
remove even this difference: it is possible to replace the inverse
determinant by the determinant of a different matrix, defining a
natural pair of mutually ``dual'' Kontsevich matrices\footnote{This
dual pair is apparently unrelated to the dual pair of boundary
cosmological constants at $c<1$.}. In terms of the dual matrix, one
then recovers a relation between correlators of determinants (rather
than inverse determinants) and the KP model.

In the concluding section we examine a peculiar property of the NMM,
namely that it describes the $c=1$ string even at finite $N$, if we
set $N=\nu$, where $\nu$ is the analytically continued cosmological
constant $\nu=-i\mu$.  This was noted in Ref.\cite{Alexandrov:2003qk},
where this variant of the NMM was called ``Model II''. Now it was
already observed in Ref.\cite{Imbimbo:1995yv} that setting $N=\nu$ in
the KP model (and giving a nonzero value to one of the deformation
parameters) reduces the KP model to the original Kontsevich model that
describes $(2,q)$ minimal strings. Thus we have a (two-step) process
leading from the NMM to the Kontsevich model. However, we also know
from Ref.\cite{Maldacena:2004sn} that the Kontsevich model arises from
insertion of macroscopic loops in the double-scaled Gaussian matrix
model. We will attempt to examine to what extent these two facts are
related.

\section{Normal Matrix Model}
\label{NMM}

We start by describing the Normal Matrix Model (NMM) of $c=1$ string
theory\cite{Alexandrov:2003qk} and making a number of observations
about it. The model originates from some well-known considerations in
the Matrix Quantum Mechanics (MQM) description of the Euclidean $c=1$
string at radius $R$. Here, $R=1$ is the selfdual radius, to which we
will specialise later. The MQM theory has discrete ``tachyons'' $T_k$,
of momentum ${k\over R}$, where $k\in Z$. Let us divide this set into
``positive tachyons'' $T_k, k>0$ and ``negative tachyons''
$T_k,k<0$. (The zero-momentum tachyon is the cosmological operator and
is treated separately). We now perturb the MQM by these tachyons,
using coupling constants $t_k,k>0$ for the positive tachyons and
$\tbar_k,k>0$ for the negative ones.

The grand canonical partition function of MQM is denoted
$\cZ(\mu,t_k,\tbar_k)$. At $t_k=\tbar_k=0$, it can easily be shown to
be:
\beq
\cZ(\mu,t_k=0,\tbar_k=0) = \prod_{n\ge 0}^\infty
\Gamma\left(- {n+\half\over R} - i\mu +\half\right)
\eeq
But this is also the partitition function of the matrix integral:
\bea
\label{ZNMMunp}
\nn {\cal Z}_{NMM} &=& \int [dZ dZ^\dag]\,e^{-\tr W(Z,Z^\dag)}\\
&=& \int [dZ dZ^\dag]\,e^{\tr\left(
-\n (ZZ^\dag)^R +\left[\half(R-1)+(R\n - N)\right] \log ZZ^\dag\right)}
\eea
where $\nu=-i\mu$ and $N\to \infty$. Here $Z$, $Z^\dag$ are
$N\times N$ matrices satisfying:
\beq
[Z,Z^\dag]=0
\eeq
Since the matrix $Z$ commutes with its adjoint, the model defined by
\eref{ZNMMunp} is called the Normal Matrix Model (NMM)\footnote{For
the most part we follow the conventions of
Ref.\cite{Alexandrov:2003qk}. However we use the transcription
$(1/i\hbar)_{\rm them}\to \nu_{\rm us}$ and $\mu_{\rm them}\to 1_{\rm
us}$. The partition function depends on the ratio $(\mu/i\hbar)_{them}\to
\nu_{us}=-i\mu_{\rm us}$. Our conventions for the NMM will be seen to
match with the conventions of Ref.\cite{Imbimbo:1995yv} for the KP
model. Note that the integral is well-defined for all complex $\nu$
with a sufficiently large real part. It can then be extended by
analytic continuation to all complex values of the parameter $\nu$,
other than those for which the argument of the $\Gamma$ function is a
negative integer. This is sufficient, since everything is ultimately
evaluated at purely imaginary values of $\nu$.}.

The equality above says that the unperturbed MQM and NMM theories are
equivalent. The final step is to note that the tachyon perturbations
correspond to infinitely many Toda ``times'' in the MQM partition
function, which becomes a $\tau$-function of the Toda integrable
hierarchy. The same perturbations on the NMM side are obtained by
adding to the matrix action the terms:
\beq
\label{NMMPert}
W(Z,Z^\dag)\to  W(Z,Z^\dag) + \nu\sum_{k=1}^\infty \left(t_k Z^k 
+ \tbar_k {Z^\dag}^k\right) 
\eeq
It follows that the Normal Matrix Model, even after perturbations, is
equivalent to MQM.

The equivalence of the full perturbed MQM and NMM gives an interesting
interpretation of the perturbations in NMM in terms of the Fermi surface of
the MQM. The unperturbed MQM Hamiltonian is given by:
\beq
\label{MQM}
H_0=\half \tr(-\hbar^2 {\del^2 \over \del X^2}-X^2)
\eeq
where $X$ is an $N\times N$ Hermitian matrix (here the compactification radius
is R). In the $SU(N)$-singlet sector this system is described by $N$
non-relativistic fermions moving in an inverted harmonic oscillator potential.
The eigenvalues of $X$ describe the positions of these fermions. In terms of
eigenvalues the Hamiltonian can be written as:
\beq
\label{MQMN}
H_0=\half \sum_{i=1}^N (\hat{p}_i^2 - \hat{x}_i^2),
\eeq
$p_i$ being the momenta conjugate to $x_i$. We now want to consider
perturbations of \eref{MQMN} by tachyon operators. For this it is convenient
to change variables from $\hat{p}$, $\hat{x}$ to the ``light cone'' variables
$\hat{x}_\pm$:
\beq
\label{lcone}
\hat{x}_\pm={\hat{x}\pm\hat{p} \over \sqrt{2}}
\eeq
Since $[\hat{p},\hat{x}]=-i\hbar$ it follows that
$[\hat{x}_+,\hat{x}_-]=-i\hbar$ also. The MQM Hamiltonian in terms of the new
variables is:
\beq
\label{MQMlcone}
H_0=-\sum_{i=1}^N\hat{x}_{+i}\hat{x}_{-i}-{i\hbar N \over 2}
\eeq
In the phase space $(x_+,x_-)$ the equation of the Fermi surface for the
unperturbed MQM is given by:
\beq
\label{Fermi0}
x_+ x_-=\mu
\eeq
The tachyon perturbations to the MQM Hamiltonian $H_0$ are given in terms of
the new variables by:
\beq
\label{HTachyon}
H=H_0-\sum_{k\geq 1}\sum_{i=1}^N\left(k\,t_{\pm k}\,x_{\pm i}^{k \over R} + 
v_{\pm k}\,x_{\pm i}^{-{k \over R}}\right)
\eeq
In the above equation the $v$'s are determined in terms of the $t$'s from the
orthonormality of the Fermion wavefunctions. The conventions chosen above
simplifies the connection with NMM perturbations. The Fermi surface of the
perturbed MQM is given by:
\beq
\label{Fermi}
x_+ x_- = \mu + \sum_{k\geq 1}\left(k\,t_{\pm k}\,x_\pm^{k \over R} + 
v_{\pm k}\,x_\pm^{-{k \over R}}\right)
\eeq

The equivalence between NMM and MQM relates the tachyon perturbations in
\eref{NMMPert} and \eref{HTachyon} with the following identification between
the tachyon operators of the two models:
\bea
\label{Identify}
\nn\tr X_+^{n \over R} & = & \tr Z^n \\
\nn\tr X_-^{n \over R} & = & \tr Z^{\dag n}
\eea
The coefficients $t_\pm$ are the same as $t,\tbar$ in the NMM. This means that
any tachyon perturbation in the NMM is mapped directly to a deformation of the
Fermi surface of MQM by \eref{Fermi}.

At the selfdual radius R=1, the NMM simplifies and the full perturbed
partition function can be written as:
\beq
\label{ZNMMReqone}
{\cal Z}_{NMM}(t,\tbar) = \int [dZ dZ^\dag]\,e^{\tr\left(
-\n ZZ^\dag+(\n - N){\rm\,log}\,ZZ^\dag-\n \sum_{k=1}^{\infty}
\big(t_k Z^k + \tbar_k {Z^\dag}^k\big)\right)}
\eeq

We note several properties of this model.

(i) The unperturbed part depends only on the combination $ZZ^\dag$ and
not on $Z,Z^\dag$ separately. 

(ii) The model can be reduced to eigenvalues, leading to the partition
function:
\beq
\label{ZNMMeigen}
{\cal Z}_{NMM}(t,\tbar) = \int \prod_{i=1}^N dz_i
d\zbar_i~\Delta(z)\Delta(\zbar)\,e^{\sum_{i=1}^N
\left(-\n z_i\zbar_i+(\n - N){\rm\,log}\,z_i\zbar_i-\n \sum_{k=1}^{\infty}
\big(t_k z_i^k + \tbar_k \zbar_i^k\big)\right)}
\eeq

(iii) The model is symmetric under the interchange $t_k\leftrightarrow
\tbar_k$, as can be seen by interchanging $Z$ and $Z^\dag$. In
spacetime language this symmetry amounts to the transformation $X\to
-X$ where $X$ is the Euclidean time coordinate, which interchanges
positive and negative momentum tachyons. 

(iv) The correlator:
\beq
\langle \tr Z^{k_1}\tr Z^{k_2}\cdots\tr Z^{k_m}
\tr {Z^\dag}^{\ell_1}\tr {Z^\dag}^{\ell_2}\cdots\tr {Z^\dag}^{\ell_n}
\rangle_{t_k=\tbar_k=0}
\eeq
vanishes unless
\beq
\sum_m k_m = \sum_n \ell_n
\eeq
This correlator is computed in the unperturbed theory. The
above result follows by performing the transformation: 
\beq
\label{transf}
Z\to e^{i\theta}Z
\eeq
for some arbitrary angle $\theta$. The unperturbed theory is invariant
under this transformation, therefore correlators that are not
invariant must vanish. In spacetime language this amounts to the fact
that tachyon momentum in the $X$ direction is conserved.

(v) As a corollary, we see that if we set all $t_k=0$, the partition
function becomes independent of $\tbar_k$:
\beq
\label{corol}
\cZ_{NMM}(0,\tbar_k) = \cZ_{NMM}(0,0)
\eeq

(vi) For computing correlators of a finite number of tachyons, it is
enough to turn on a {\it finite} number of $t_k,\tbar_k$, i.e. we can
always assume for such purposes that $t_k,\tbar_k=0, k>k_{max}$ for
some finite integer $k_{max}$. In that case, apart from the $\log$
term we have a polynomial matrix model.

(vii) We can tune away the $\log$ term by choosing $\nu=N$. This
choice has been called Model II in Ref.\cite{Alexandrov:2003qk}. In
this case the model reduces to a Gaussian model (but of a normal,
rather than Hermitian, matrix) with perturbations that are holomorphic
+ antiholomorphic in the matrix $Z$ (i.e., in the eigenvalues
$z_i$). If we assume that the couplings $t_k,\tbar_k$ vanish for
$k>k_{max}$, as in the previous comment, then the perturbations are
also polynomial. We will return to this case in a subsequent section.

\section{The Kontsevich-Penner or $W_\infty$ model}

The Kontsevich-Penner or $W_\infty$ model\cite{Imbimbo:1995yv} (for a
more detailed review, see Ref.\cite{Mukhi:2003sz}) is a model of a
single positive-definite hermitian matrix, whose partition function is
given by:
\beq
\label{ZKP}
{\cal Z}_{KP}(A,\tbar) = (\det A)^\n \int [dM]\, e^{\tr\left(
-\n MA+(\n - N){\rm\,log}\,M-\n \sum_{k=1}^{\infty}
\tbar_k M^k \right)}
\eeq
where $\tbar_k$ are the couplings to negative-momentum
tachyons, $N$ is the dimensionality of the matrix $M$ and $A$ is a
constant matrix. The eigenvalues of this matrix determine
the couplings $t_k$ to positive-momentum tachyons via the
Kontsevich-Miwa (KM) transform:
\beq 
\label{TK}
t_k = -\frac{1}{\n k} \tr(A^{-k}) 
\eeq

This model is derived by integrating the $W_\infty$ equations found in
Ref.\cite{Dijkgraaf:1992hk}. The parameter $\nu$ appearing in the
action above is related to the cosmological constant $\mu$ of the
string theory by $\nu=-i\mu$. The model can also be obtained from the
Penner matrix model\cite{Penner,DisVaf} after making a suitable change
of variables (as explained in detail in Ref.\cite{Mukhi:2003sz}) and
adding perturbations.

We now note some properties that are analogous to those of the NMM, as
well as others that are quite different.

(i) By redefining $MA\to M$ we can rewrite the partition function
without any factor in front, as:
\beq
\label{ZKPalt}
{\cal Z}_{KP}(A,\tbar) = \int [dM]\, e^{\tr\left(
-\n M+(\n - N){\rm\,log}\,M-\n \sum_{k=1}^{\infty}
\tbar_k (MA^{-1})^k \right)}
\eeq

(ii) This model has no radius deformation, and describes the $c=1$ string
theory directly at selfdual radius $R=1$.

(iii) In view of the logarithmic term, the model is well-defined only if
the integral over the eigenvalues $m_i$ of the matrix $M$ is
restricted to the region $m_i>0$. 

(iv) The model can be reduced to eigenvalues, leading to the partition
function:
\beq
\label{kpeigen}
{\cal Z}_{KP}(A,\tbar) = \left(\prod_{i=1}^N a_i\right)^\nu
\int \prod_{i=1}^N dm_i~
{\Delta(m)\over \Delta(a)}\,
e^{\sum_{i=1}^N\left(-\n m_i a_i+(\n - N){\rm\,log}\,m_i-\n \sum_{k=1}^{\infty}
\tbar_k m_i^k \right)}
\eeq

(vi) In the representation \eref{ZKP}, the operators $\tr M^k$
describe the negative-momentum tachyons. But there are no simple
operators that directly correspond to positive-momentum
tachyons. Nevertheless this model generates tachyon correlators of the
$c=1$ string as follows:
\beq
\langle \cT_{k_1}\cT_{k_2}\cdots \cT_{k_m}
T_{-\ell_1}T_{-\ell_2}\cdots T_{-\ell_n}\rangle
= {\del\over\del t_{k_1}}{\del\over\del t_{k_2}}\cdots{\del\over\del t_{k_m}}
{\del\over\del \tbar_{\ell_1}}{\del\over\del \tbar_{\ell_2}}\cdots
{\del\over\del \tbar_{\ell_n}}\log\cZ_{KP}
\eeq
where derivatives in $t_k$ are computed using \eref{TK} and the chain
rule.

(v) The symmetry of the partition function under the interchange of 
$t_k,\tbar_k$ is not manifest, since one set of parameters is encoded
through the matrix $A$ while the other appears explicitly.

(vi) The transformation 
\beq
A\to \alpha A,\qquad \tbar_k\to \alpha^k\,\tbar_k
\eeq
for arbitrary $\alpha$, is a symmetry of the model (most obvious
in the representation \eref{ZKPalt}). As a consequence, the tachyon correlators
satisfy momentum conservation.

(vi) The partition function satisfies the ``puncture equation'':
\beq
\cZ_{KP}(A-\epsilon, \tbar_k + \delta_{k,1}\,\epsilon) = e^{\epsilon\nu^2 t_1}
\cZ_{KP}(A,\tbar_k)
\eeq
as can immediately be seen from \eref{ZKP}.

\section{Equivalence of the matrix models}
\label{Equivalence}

\subsection{$N=1$ case}

We start by choosing the selfdual radius $R=1$, and will later comment
on what happens at other values of $R$. As we have seen, in the
perturbed NMM there are two (infinite) sets of parameters
$t_k,\tbar_k$, all of which can be chosen independently. This is the
case even at finite $N$, though the model describes $c=1$ string
theory only at infinite $N$ (or at the special value $N=\nu$, as noted
in Ref.\cite{Alexandrov:2003qk}, a point to which we will return
later). In contrast, the Kontsevich-Penner model has one infinite set of
parameters $\tbar_k$, as well as $N$ additional parameters from the
eigenvalues of the matrix $A$. The latter encode the $t_k$, as seen
from \eref{TK} above. From this it is clear that at finite $N$, there
can only be $N$ independent parameters $t_k$ ($k=1,2,\ldots N$) while
the remaining ones ($t_k, k>N$) are dependent on these.

This makes the possible equivalence of the two models somewhat subtle. To
understand the situation better, let us compare both models in the
limit that is farthest away from $N\to\infty$, namely $N=1$. While
this is a ``toy'' example, we will see that it provides some useful
lessons. 

In this case the NMM partition function is:
\beq
\cZ_{NMM,N=1}(t_k,\tbar_k) = \int dz\, d\zbar~ e^{-\nu z\zbar + (\nu-1)\log z\zbar
-\nu\sum_{k=1}^\infty(t_kz^k + \tbar_k\zbar^k)}
\eeq
while the Kontsevich-Penner partition function is:
\beq
\cZ_{KP, N=1}(a,\tbar_k) = a^\nu\int dm~ e^{-\nu ma + (\nu-1)\log m
-\nu\sum_{k=1}^\infty \tbar_k m^k}
\eeq
We will now show that the two integrals above are equivalent if we
assume that $t_k$ in the NMM is given by:
\beq
\label{kmone}
t_k = -{1\over \nu k}\,a^{-k}
\eeq
which is the KM transform \eref{TK} in the special case
where $A$ is a $1\times 1$ matrix, denoted by the single real number
$a$.  Note that this determines all the infinitely many $t_k$ in terms
of $a$.

To obtain the equivalence, insert the above relation and also perform the
change of integration variable:
\beq
z = \sqrt{m}\,e^{i\theta}
\eeq
in the NMM integral. Then we find that (up to a numerical constant):
\bea
\nn
\cZ_{NMM,N=1}(a,\tbar_k) &=& \int dm\, d\theta~ e^{-\nu m + (\nu-1)\log m
+\sum_{k=1}^\infty {1\over k} ({\sqrt{m}\over a})^k e^{ik\theta}
-\nu\sum_{k=1}^\infty \tbar_k (\sqrt{m})^k e^{-ik\theta}}\\
&=& \int dm\, d\theta~ {1\over 1-{\sqrt{m}\,e^{i\theta}\over a}}\,e^{-\nu m + (\nu-1)\log m
-\nu\sum_{k=1}^\infty \tbar_k (\sqrt{m})^k e^{-ik\theta}}
\eea
Strictly speaking the last step is only valid for $\sqrt{m}/a<1$, 
since otherwise the infinite sum fails to converge. Hence we
fix $m$ and $a$ to satisfy this requirement and continue by evaluating
the $\theta$-integral. This can be evaluated by defining $e^{-i\theta}
= w$ and treating it as a contour integral in $w$. We have
\beq
d\theta\, {1\over 1-{\sqrt{m}\,e^{i\theta}\over a}}\to dw\,{1\over
w-{\sqrt{m}\over a}}
\eeq
Since the rest of the integrand is well-defined and analytic near
$w=0$, we capture the simple pole at $w=\sqrt{m}/a$. That
brings the integrand to the desired form. Now we can lift the 
restriction $\sqrt{m}/a<1$, and treat the
result as valid for all $m$ by analytic continuation. Therefore
we find:
\bea
\nn
\cZ_{NMM,N=1}(a,\tbar_k) &=& \int dm~e^{-\nu m + (\nu-1)\log m
-\nu\sum_{k=1}^\infty \tbar_k (m a^{-1})^k}\\
&=& \nn a^\nu \int dm~e^{-\nu ma + (\nu-1)\log m
-\nu\sum_{k=1}^\infty \tbar_k m^k}\\
&=& \cZ_{KP, N=1}(a,\tbar_k)
\eea
Thus we have shown that the perturbed $1\times 1$ Normal Matrix Model
at $R=1$ is equivalent to the $1\times 1$ Kontsevich-Penner
model. However, this equivalence only holds when we perform the
$1\times 1$ KM transform, which fixes all the perturbations $t_k$ in
terms of a single independent parameter $a$ (while the $\tbar_k$ are
left arbitrary).

An important point to note here is the sign chosen in
\eref{kmone}. Changing the sign (independently of $k$) amounts to 
the transformation $t_k\to -t_k$. This is apparently harmless, leading
to some sign changes in the correlation functions, but there is no way
at $N=1$ (or more generally at any finite $N$) to change $a$ (or the
corresponding matrix $A$) to compensate for this transformation. The
sign we have chosen, given the signs in the original NMM action, is
therefore the only one that gives the KP model. This point will become
important later on.

Returning to the NMM-KP equivalence at $N=1$, it is interesting to
generalise it by starting with the NMM at an arbitrary radius $R$
instead of $R=1$ as was the case above. As seen from \eref{ZNMMunp},
the coupling of the log term is modified in this case as:
\beq
(\nu -1) \to \half(R-1) + (R\nu -1)
\eeq
and also the bilinear term $z\zbar$ is modified to $(z\zbar)^R$.
The above derivation goes through with only minor changes, and we end
up with:
\beq
\cZ_{NMM,N=1}(a,\tbar_k) = a^{\half(R-1)+\nu} \int dm~e^{-\nu (ma)^R + 
\left[\half(R-1)+(R\nu-1)\right]\log m
-\nu\sum_{k=1}^\infty \tbar_k m^k}
\eeq
This appears to suggest that there is a variant of the
Kontsevich-Penner model valid at arbitrary radius (or at least
arbitrary integer radius, since otherwise it may become hard to define
the integral). This would be
somewhat surprising as such a model has not been found in the past. As
we will see in the following subsection, the above result holds only
for the $N=1$ case. Once we go to $N\times N$ matrices, we will see
that NMM leads to a KP matrix model only at $R=1$, consistent with
expectations.

Another generalisation of the above equivalence seems more
interesting. In principle, even for the $1\times 1$ matrix model, we
can carry out a KM transform using an $n\times n$ matrix $A$ where $n$
is an arbitrary integer. Indeed, there is no logical reason why the
dimension of the constant matrix $A$ must be the same as that of the
random matrices occurring in the integral. The most general example of
this is to take $N\times N$ random matrices $Z,Z^\dag$ in the NMM and
then carry out a KM transform with $A$ being an $n\times n$
matrix. The ``usual'' transform then emerges as the special case
$n=N$. Of course all this makes sense only within the NMM and not in
the KP model. If $n\ne N$ then the KP model, which has a $\tr MA$ term
in its action, cannot even be defined. So we should not expect to find
the KP model starting with the NMM unless $n=N$, but it is still
interesting to see what we will find.

Here we will see what happens if we take $N=1$ and $n>1$. The
full story will appear in a later subsection.  Clearly the KM
transform \eref{TK} permits more independent parameters $t_k$ as $n$
gets larger. Let us take the eigenvalues of $A$ to be
$a_1,a_2,\ldots,a_n$. Then it is easy to see that:
\bea
\nn
\cZ_{NMM,N=1}(a_i,\tbar_k) &=& \int dm\, d\theta~ e^{-\nu m + (\nu-1)\log m
+\sum_{i=1}^n\sum_{k=1}^\infty {1\over k} ({\sqrt{m}\over a_i})^k e^{ik\theta}
-\nu\sum_{k=1}^\infty \tbar_k (\sqrt{m})^k e^{-ik\theta}}\nonumber\\
&=& \int dm\, d\theta~ {1\over
\prod_{i=1}^n(1-{\sqrt{m}\,e^{i\theta}\over a_i})}\,
e^{-\nu m + (\nu-1)\log m
-\nu\sum_{k=1}^\infty \tbar_k (\sqrt{m})^k e^{-ik\theta}}
\eea
Converting to the $w$ variable as before, we now encounter $n$
poles. Picking up the residues, we get:
\beq
\cZ_{NMM,N=1}(a_i,\tbar_k) = \int dm~  e^{-\nu m + (\nu-1)\log m}
\sum_{l=1}^n\left({1\over \prod_{i\ne l}\big(1-{a_l\over a_i}\big)}
e^{-\nu\sum_{k=1}^\infty \tbar_k ({m\over a_l})^k}\right)
\eeq
This in turn can be expressed as a sum over $n$ $1\times 1$
Kontsevich-Penner models:
\beq
\label{sumoverKP}
\cZ_{NMM,N=1}(a_i,\tbar_k) = \sum_{l=1}^n{1\over \prod_{i\ne
l}\big(1-{a_l\over a_i}\big)} \cZ_{KP,N=1}(a_l,\tbar_k)
\eeq
Note that if in this expression we take $a_n\to\infty$, one of the
terms in the above equation (corresponding to $l=n$) decouples, and
$a_n$ also drops out from the remaining terms. Therefore we recover
the same equation with $n\to n-1$. In this way we can successively
decouple all but one of the $a_i$'s.

To summarise, at the level of the $1\times 1$ NMM, we have learned some
interesting things: this model is equivalent to the $1\times 1$ KP
model if we specialise the parameters $t_k$ to a 1-parameter family
via the KM transform, while it is equivalent to a sum
over $n$ different $1\times 1$ KP models if we specialise the
parameters $t_k$ to an $n$-parameter family. We also saw a $1\times 1$
KP model arise when we are at a finite radius $R\ne 1$.  In the next
section we will see to what extent these lessons hold once we work
with $N\times N$ random matrices.

\subsection{General case}

In this section we return to the $N\times N$ Normal Matrix Model. With
the substitution \eref{TK} (where $A$ is also an $N\times N$ matrix),
its partition function becomes:
\beq
\label{ZNMOD}
{\cal Z}_{NMM} = \int [dZ\, dZ^\dag]\,e^{\tr\left(\
-\n ZZ^\dag+(\n - N){\rm\,log}\,ZZ^\dag+\sum_{k=1}^{\infty}
\frac{1}{k} \tr(A^{-k}) Z^k
-\n \sum_{k=1}^{\infty} \overline{t}_k {Z^\dag}^k\right)}
\eeq
or, in terms of eigenvalues:
\bea
\label{ZNEIG}
\nn {\cal Z}_{NMM}  &=&  \int \prod_{i=1}^N d^2z_i~ 
\Delta(z)\Delta(\zbar)\,
e^{-\n\sum_{i=1}^N z_i\overline{z}_i+(\n - N) \sum_{i=1}^N {\rm log}\,
z_i\overline{z}_i} \\
& &\times~ e^{\sum_{i,j=1}^N \sum_{k=1}^{\infty} \frac{1}{k}
 \left(\frac{z_i}{a_j}\right)^k}
 e^{-\n \sum_{i=1}^N \sum_{k=1}^{\infty} \overline{t}_k
 \overline{z}_i^k}
\eea
where $\Delta(z)$ is the Vandermonde determinant. Because of the
normality constraint $[Z,Z^\dag]=0$ there is only one Vandermonde for
$z_i$ and one for $\zbar_i$.

The sum over $k$ in the second line of \eref{ZNEIG} converges if
${z_i\over a_j}<1$ for all $i,j$, in which case it can be evaluated immediately
giving:
\beq
\label{ZNEIG2}
\nn {\cal Z}_{NMM}  =  \int \prod_{i=1}^N d^2z_i~ |\Delta(z)|^2\,
\prod_{i,j=1}^N \frac{1}{1 - \frac{z_i}{a_j}}~
e^{\sum_{i=1}^N\left[-\n z_i\zbar_i+(\n - N) {\rm log}\,
z_i\zbar_i -\n \sum_{k=1}^{\infty} \overline{t}_k
\zbar_i^k\right]}
\eeq

To make contact with the Penner model, first change variables $z_i \ra
\sqrt{m_i}\,e^{i\te_i}$ and then replace $e^{-i\theta_i}$ by $w_i$ as
before. Then we get $d^2z_i \ra dm_i\, {dw_i\over w_i}$ and:
\bea
\label{ZNEIG3}
\nn {\cal Z}_{NMM} & = & \int \prod_{i=1}^N dm_i\oint \prod_{i=1}^N
{dw_i\over w_i}~
\prod_{i<j}^N \left({\sqrt{m_i}\over w_i} - {\sqrt{m_j}\over
w_j}\right)\left(\sqrt{m_i} w_i - \sqrt{m_j} w_j\right)\\
&&\times\,\prod_{i,j=1}^N \frac{1}{1 - \frac{\sqrt{m_i}}{w_i a_j}}
~e^{\sum_{i=1}^N\left[-\n m_i+(\n - N) \log m_i 
-\n \sum_{k=1}^{\infty} \tbar_k (\sqrt{m_i}w_i)^k  \right]}
\eea
The contour integrals can be evaluated once this is rewritten in the
more convenient form: 
\bea
\label{ZNCONT}
\nn {\cal Z}_{NMM} & = & \int \prod_{i=1}^N dm_i\oint 
\prod_{i=1}^N dw_i~
\prod_{i<j}^N \left(\sqrt{m_i} w_j - \sqrt{m_j} w_i\right)
\left(\sqrt{m_i} w_i - \sqrt{m_j} w_j\right)\\
&&\times\,\prod_{i,j=1}^N \frac{1}{w_i - \frac{\sqrt{m_i}}{a_j}}
~e^{\sum_{i=1}^N\left[-\n m_i+(\n - N) \log m_i 
-\n \sum_{k=1}^{\infty} \tbar_k (\sqrt{m_i}w_i)^k  \right]}
\eea

Next we pick up the residues at the poles. During the intermediate
steps, we will assume that the eigenvalues $a_i$ of the matrix $A$ are
non-degenerate. From the above expression, each integration variable
$w_i$ has a pole at each of the values:
\beq
w_i = {\sqrt{m_i}\over a_j}
\eeq
for all $j$. Thus the contributions can be classified by the set of
poles:
\beq
\label{poleclassif}
(w_1,w_2,\ldots, w_N) = \left({\sqrt{m_1}\over a_{j_1}},{\sqrt{m_2}\over
a_{j_2}},\ldots, {\sqrt{m_N}\over a_{j_N}}\right)
\eeq
We now notice that the set $(j_1,j_2,\ldots,j_N)$ must consist of
distinct elements, in other words it forms a permutation of
$(1,2,\ldots,N)$. This is because if two values of $j_i$ coincide, one
of the Vandermonde factors of the type $(\sqrt{m_i} w_j -
\sqrt{m_j}w_i)$ vanishes and there is no contribution.

We start by considering the simplest permutation, the identity, namely:
\beq
(j_i,j_2,\ldots,j_N)=(1,2,\ldots,N)
\eeq
In this case the residues from the denominator and Vandermonde factors
become:
\beq
\prod_{i<j}^N \left({\sqrt{m_i m_j}\over a_j} - {\sqrt{m_i m_j}\over a_i}\right)
\left({m_i\over a_i} - {m_j\over a_j}\right)\,
\prod_{j\ne i}^N \frac{1}{{\sqrt{m_i}\over a_i} - \frac{\sqrt{m_i}}{a_j}}
= {\prod_{i<j}^N(m_ia_j-m_ja_i)\over \Delta(a)}
\eeq
while the exponential measure factor becomes:
\beq
e^{\sum_{i=1}^N\left[-\n m_i+(\n - N) \log m_i 
-\n \sum_{k=1}^{\infty} \tbar_k \left({m_i\over a_i}\right)^k  \right]}
\eeq
It is easy to check that for all the other possible permutations of
$(j_1,j_2,\ldots,j_N)$ besides the identity permutation, a
corresponding permutation of the integration variables $m_i$ brings
the above answer (exponential measure as well as prefactors) back to
the same form as for the identity permutation. This means that
(dropping a factor of ${1\over N!}$) we have proved:
\bea
\label{evalrep}
\nn \cZ_{NMM} &=& \int \prod_{i=1}^N dm_i~
{\prod_{i<j}^N(m_ia_j-m_ja_i)\over \Delta(a)}
e^{\sum_{i=1}^N\left[-\n m_i+(\n - N) \log m_i 
-\n \sum_{k=1}^{\infty} \tbar_k \left({m_i\over a_i}\right)^k
\right]}\\
&=& \Big(\prod_{i=1}^N a_i\Big)^\nu \int   \prod_{i=1}^N dm_i~
{\Delta(m)\over \Delta(a)}\,
e^{\sum_{i=1}^N\left[-\n m_i a_i+(\n - N) \log m_i 
-\n \sum_{k=1}^{\infty} \tbar_k m_i^k
\right]}
\eea
where in the last step we have replaced $m_i \to m_i a_i$.

This is precisely the eigenvalue representation \eref{kpeigen} of the
KP matrix model \eref{ZKP}. Thus we have provided a direct proof of
equivalence of the perturbed Normal Matrix Model and the
Kontsevich-Penner model. Notice that in performing the KM transform we
reduced the independent $t_k$ of the NMM to a finite number, namely
$N$, so that eventually the $N\to\infty$ limit is required in order to
encode all the independent parameters.

In the previous subsection we considered taking different ranks for
the constant matrix $A$ arising in the KM transform and the random
matrix $Z$. The most general case is to take $Z$ to be $N\times N$ and
$A$ to be $n\times n$. The computation is a simple extension of the
one done above. We find the following results. When $n>N$ we again get
a sum over Kontsevich-Penner models. The number of terms in the sum is
the binomial coefficient ${}^nC_N$. This is a generalisation of the
result given in \eref{sumoverKP} for $N=1$, where we found $n$
terms. In the general case let us denote by $a_{\{i,l\}}$ the $i^{\rm
th}$ element of the set formed by one possible choice of $N$ $a_i$'s
from a total of $n$, the index $l$ labeling the particular choice. The
complementary set, formed by the rest of the $a_i$'s is denoted by
$a_{\{\tilde{i},l\}}$, the index $\tilde{i}$ taking $n-N$ values.  We
then have:
\beq
\label{sumoverKPN}
\cZ_{NMM}(a_i,\tbar_k) = \sum_{l=1}^{{}^n C_N}  
\prod_{i=1}^N \prod_{\tilde{i}=1}^{N-n} 
{1 \over \left(1 - {a_{\{i,l\}} \over a_{\{\tilde{i},l\}}}\right)}
\cZ_{KP}(a_{\{l\}},\tbar_k)
\eeq
so that the NMM is again expressed as a sum over KP models.

The other case, $n<N$, can be obtained by starting with $n=N$ and
successively decoupling $N-n$ eigenvalues $a_i$ by taking them to
infinity. This is similar to what we observed in the $N=1$ case
following \eref{sumoverKP}. In the present case one can easily show
that $N-n$ matrix eigenvalues $m_i$ also decouple in this limit (apart
from a normalisation). In fact, it is straightforward to derive the
formula:
\beq
\lim_{a_N\to\infty} Z_{KP}^{(N,\nu)}(A^{(N)},\tbar_k)
= {\Gamma(\nu-N+1)\over \nu^{\nu-N+1}} Z_{KP}^{(N-1,\nu)}
(A^{(N-1)},\tbar_k)
\eeq
which can then be iterated. Thus after $N-n$ eigenvalues $a_i$ are
decoupled, we find up to normalisation the KP model of rank $n$. As we
remarked in the introduction, this exhibits a strong analogy to the
insertion of $n$ determinant operators in the Gaussian model, as
described in Ref.\cite{Maldacena:2004sn}, where the result is the
$n\times n$ Kontsevich model\footnote{This paragraph corrects an error
in a previous version of this paper. As a result, the analogy with
$c<1$ is now {\it stronger} than we had previously claimed. We
are grateful to the referee for helpful suggestions in this regard.}.

\subsection{Radius dependence}

Finally, we can ask what happens to the radius-dependent NMM under the
above procedure. Again the steps are quite straightforward and one
arrives at the following generalisation of \eref{evalrep}:
\bea
\label{evalmod}
\nn\cZ_{NMM,R} &=& \Big(\prod_{i=1}^N a_i\Big)^{\half(R-1)+\nu} 
\int  \prod_{i=1}^N dm_i~
{\Delta(m)\over \Delta(a)}\\
&&\times\,
e^{\sum_{i=1}^N\left[-\n (m_i a_i)^R+\left[\half(R-1)+(\n R - N)\right] \log m_i 
-\n \sum_{k=1}^{\infty} \tbar_k m_i^k
\right]}
\eea
The problem is that the above eigenvalue model cannot (as far as we
can see) be converted back to a matrix model. The key to doing so in
the $R=1$ case was the linear term $\sum_i m_ia_i$ in the action,
which (after absorbing the Vandermondes and using the inverse of the 
famous Harish Chandra formula) can be summed
back into $\tr MA$. The quantity $\sum_i(m_ia_i)^R$
cannot be converted back into a matrix trace unless $R=1$.

This clarifies a longstanding puzzle: while a KP model could only be
found at $R=1$, the NMM exists and describes the $c=1$ string for any
$R$. We see now that the correct extension of the KP model to $R\ne 1$ is the
eigenvalue model given by \eref{evalmod} above, but unfortunately this
does not correspond to a matrix model.

\section{Loop operators in the NMM}
\label{loopops}

In this section we will examine loop operators in the NMM. Our goal
here is to understand whether correlation functions of these operators
can be related to the Kontsevich-Penner model of
Ref.\cite{Imbimbo:1995yv}, thereby providing the $c=1$ analogue of the
corresponding observations in
Refs.\cite{Maldacena:2004sn,Hashimoto:2005bf}. Though there are some
similarities, we will also find some striking differences between this
and the $c<1$ case.

Macroscopic loops in a model of random matrices $\Phi$ are described by
insertions of the operator:
\beq
\label{macloop}
W(x) = \tr \log (x-\Phi)
\eeq
which creates a boundary in the world sheet. Here $x$ is the boundary
cosmological constant. The corresponding generating function for
multiple boundaries is\cite{Banks:1989df,Moore:1991ag,Kostov:1991hn,%
Kutasov:2004fg,Maldacena:2004sn}:
\beq
\label{multipleloop}
e^{W(x)} = \det(x-\Phi)
\eeq
Such operators have been studied extensively in $c<1$ matrix models,
describing $(p,q)$ minimal models coupled to 2d gravity.

We will consider expectation values of operators of the form $\det(a
- Z)$ in the NMM, where $a$ is a real parameter. These operators
create a hole in the dual graph in the Feynman diagram expansion of
the matrix model. Since the NMM has vertices that are
holomorphic/antiholomorphic in $Z$, the dual graph will have faces
that are dual to $Z$ or $Z^\dag$. The loop operator $\det(a- Z)$
creates a hole in a $Z$-face, while its complex conjugate creates a
hole in a $Z^\dag$-face.

As we would expect, this means that the correlators are complex, 
but we have the identity\footnote{Here and in the rest of this
section, all correlators are understood to be normalised correlators
in the NMM.}:
\beq
\blangle\prod_i\det(a_i-Z)\brangle_{t_k,\tbar_k} = 
\blangle\prod_i\det(a_i-Z^\dag)\brangle_{\tbar_k,t_k}
\eeq
where on the RHS the role of the deformations $t_k,\tbar_k$ has been
interchanged. Therefore as long as we consider correlators only of
$\det(a_i-Z)$ or $\det(a_i-Z^\dag)$ the result is effectively the
same. As we will see in a moment, a stronger statement is true: on the
subspace of parameter space dictated by the KM transform, the unmixed
correlators are individually real. Later we will also consider mixed correlators.

As a start, notice that in the $1\times 1$ case,
\bea
\nn\cZ_{NMM,N=1}(t_k=0,\tbar_k) &=&
\int d^2z~ e^{-\nu z\zbar + (\nu-1)\log z\zbar
-\nu\sum_{k=1}^\infty \tbar_k\zbar^k}\\
\nn&=& \int d^2z~ (a-z){1\over (a-z)}\,e^{-\nu z\zbar + (\nu-1)\log z\zbar
-\nu\sum_{k=1}^\infty \tbar_k\zbar^k}\\
\nn&=& {1\over a}\int d^2z~ (a-z)\,e^{-\nu z\zbar + (\nu-1)\log z\zbar
-\nu\sum_{k=1}^\infty(t^0_k z^k +  \tbar_k\zbar^k)}\\
&=& {1\over a}\bigg\langle(a-z)\bigg\rangle_{t^0_k,\tbar_k}
~Z_{NMM,N=1}(t^0_k,\tbar_k)
\eea
where the expectation value in the last line is evaluated in the NMM with
\beq
\label{tzero}
t^0_k = -{1\over\nu k}\,a^{-k}
\eeq
We see that the $t^0_k$
dependence drops out in the RHS because insertion of the loop
operator cancels the dependence in the partition function. In fact,
more is true: even the $\tbar_k$ dependence cancels out between the
different factors on the RHS. This is a consequence of the property
exhibited in \eref{corol}.

A more general statement in the $1\times 1$ case is:
\beq
\nn\cZ_{NMM,N=1}(t_k-t^0_k,\tbar_k) 
= {1\over a}\bigg\langle(a-z)\bigg\rangle_{t_k,\tbar_k}~Z_{NMM,N=1}(t_k,\tbar_k)
\eeq
In other words, insertion of the macroscopic loop operator has the
effect of decreasing the value of $t_k$, leaving $\tbar_k$ unchanged. 

In the more general case of $N\times N$ random matrices, the
corresponding result is as follows. The expectation value of a single
exponentiated loop operator $\det (a-Z)$ is:
\beq
\blangle \det(a-Z)\brangle_{t_k,\tbar_k} = 
{\cZ_{NMM}(t_k-t^0_k,\tbar_k) \over
\cZ_{NMM}(t_k,\tbar_k)}\,a^N
\eeq
with $t^0_k$ again given by \eref{tzero}.
Now we would like to consider multiple loop operators. Therefore
consider the expectation value:
\beq
\blangle\prod_{i=1}^n \det(a_i-Z)\brangle_{t_k,\tbar_k}
\eeq
As noted in Ref.\cite{Maldacena:2004sn}, this can be thought of as a
single determinant in a larger space. Define the $n\times n$ matrix
$A={\rm diag}(a_1,a_2,\ldots,a_n)$ and extend it to an
$(n+N)\times (n+N)$ matrix $A\otimes \Id_{N\times N}$.  Similarly,
extend the $N\times N$ matrix $Z$ to an $(n+N)\times (n+N)$ matrix
$\Id_{n\times n}\otimes Z$. Now we can write
\beq
\label{detcorr}
\prod_{i=1}^n \det(a_i-Z) = \det(A\otimes\Id -\Id\otimes Z)
= \prod_{i=1}^n\prod_{j=1}^N(a_i-z_j)
\eeq
Rewriting this as:
\beq
\prod_{i=1}^n \det(a_i-Z) = (\det A)^N\det(\Id\otimes\Id -A^{-1}\otimes Z)
\eeq
and expanding the second factor, we find:
\beq
\blangle \prod_{i=1}^n\det(a_i-Z)\brangle_{t_k,\tbar_k} = 
{\cZ_{NMM}(t_k-t^0_k,\tbar_k) \over
\cZ_{NMM}(t_k,\tbar_k)}\,(\det A)^N
\eeq
where now:
\beq
\label{tzeron}
t^0_k = -{1\over\nu k}\,\tr A^{-k}
\eeq
Thus we see that macroscopic loop correlators in this model are
obtained by simply shifting the parameters $t_k$ in the partition
function, the shift being given by the KM transform.

The above considerations can be extended to mixed correlators as
follows. Consider correlation functions of the form: 
\beq
\blangle \prod_{i=1}^n\det(a_i-Z)
\prod_{j=1}^m\det(b_j-Z^\dag)\brangle
\eeq
Then, defining the $m\times m$ matrix $B={\rm
diag}(b_1,b_2,\ldots,b_m)$, the parameters $t^0_k$ as in
\eref{tzeron}, and the parameters $\tbar^0_k$ by:
\beq
\label{tbarzeron}
\tbar^0_k = -{1\over\nu k}\,\tr B^{-k}
\eeq
we find
\beq
\blangle\prod_{i=1}^n\det(a_i-Z)
\prod_{j=1}^m\det(b_j-Z^\dag)\brangle_{t_k,\tbar_k}
= {\cZ_{NMM}(t_k-t^0_k,\tbar_k-\tbar^0_k) \over
\cZ_{NMM}(t_k,\tbar_k)}\,(\det A\det B)^N
\eeq

In the above we have seen how to re-express correlations of loop
operators in terms of shifted closed-string parameters. This in itself
is quite reminiscent of an open-closed duality. However we did not yet
encounter the KP model. To do so, we note that besides the
exponentiated loop operator $\det(a-Z)$, we can consider its inverse:
${1/\det(a-Z)}$. Just as insertion of $\det(a-Z)$ has the effect of
decreasing each $t_k$ by $t^0_k$ given by \eref{tzero}, insertion of
the inverse operator {\it increases} $t_k$ by the same
amount.

Thus we may consider correlators like:
\beq
\label{invdet}
\blangle\prod_{i=1}^n {1\over\det(a_i-Z)}\brangle
= {1\over(\det A)^N}
\left\langle {1\over\det(\Id\otimes\Id-A^{-1}\otimes Z)}\right\rangle
\eeq
As before, the two factors of the direct product in the above equation
refer to $n\times n$ and $N\times N$ matrices. It is easy to see that
the correlation function on the RHS has the effect of increasing the
$t_k$ by $t^0_k$ as given in \eref{tzeron}.

Although in principle $n$ and $N$ are independent, here we will
consider the case $n=N$. Now the inverse
operator
\beq
\left\langle {1\over\det(\Id\otimes\Id-A^{-1}\otimes Z)}\right\rangle
\eeq
 has already made an appearance in
\sref{Equivalence}, where one finds it in the eigenvalue basis (see
for example \eref{ZNEIG2}):
\beq
\prod_{i,j=1}^N {1\over 1-{z_i\over a_j}}
\eeq

The interesting property of the inverse determinant operators is that
they can be used to create the KP model starting from the {\it
partially unperturbed} NMM (where $t_k=0$ but $\tbar_k$ are
arbitrary). Computationally this is similar to the derivation in
\sref{Equivalence} of the KP model from the perturbed
NMM. Thus we have:
\beq
\left\langle {1\over\det(\Id\otimes\Id-A^{-1}\otimes
Z)}\right\rangle_{0,\tbar_k} \cZ_{NMM}(0,\tbar_k)
= Z_{KP}(A,\tbar_k)
\eeq
Here $\cZ_{NMM}(0,\tbar_k)$ can be replaced by $\cZ_{NMM}(0,0)$ as we
have noted previously. This equation then is the precise statement of
one of our main observations, that inverse determinant expectation
values in the (partially unperturbed) NMM give rise to the KP partition function.

It is clearly desirable to have a target space interpretation for
these loop operators. Since the NMM is derived from correlators
computed from matrix quantum mechanics, in principle one should be
able to understand the loop operators of NMM starting from loop
operators (or some other operators) in MQM. While that is beyond the
scope of the present work, we will instead exhibit some suggestive
properties of our loop operators and leave their precise interpretation
for future work.

In matrix models for the $c<1$ string, which are described by constant
random matrices, exponentiated loop operators are determinants just
like the ones discussed here for the NMM. In those models it has been
argued that the loop operators represent FZZT branes. One striking
observation is that in the Kontsevich/generalised Kontsevich
description of $c<1$ strings, the eigenvalues of the constant matrix
$A$ come from the boundary cosmological constants appearing in the
loop operators. Moreover,
\eref{multipleloop} has been interpreted as evidence that the FZZT-ZZ
open strings there are
fermionic\cite{Kutasov:2004fg,Maldacena:2004sn}. 

In the present case, we see that the parameters $a_i$ in the loop
operators turn precisely into the eigenvalues of the constant matrix
$A$ of the Kontsevich-Penner model. We take this as evidence that our
loop operators are likewise related in some way to FZZT
branes. Indeed, one is tempted to call them FZZT branes of the
NMM. Pursuing this analogy further, the role played by inverse
determinants in the present discussion appears to suggest that the
corresponding strings in the NMM are {\it bosonic} rather than
fermionic. But the relationship of these operators to the ``true''
FZZT branes of matrix quantum mechanics remains to be understood, as
we have noted above\footnote{In light of the discussions about the MQM
Fermi surface in \sref{NMM} we can give an interpretation to both
determinant and inverse determinant operators in the NMM. Since their
insertions lead to opposite shifts in the $t_k$'s, by virtue of the
equivalence between MQM and NMM discussed above we can map each one
directly to a corresponding deformation to the Fermi surface, which
can be read off from \eref{Fermi}. This fact should facilitate direct
comparison with the MQM.}.

In the limit of infinite $N$, the inverse loop operators depending on
a matrix $A$ can be thought of as loop operators for a different
matrix $\tA$. Thus, only in this limit, the inverse determinant
operators can be replaced by more conventional determinants. This
proceeds as follows. We have already seen that the KM transformation
\eref{TK} encodes infinitely many parameters $t_k$ via a constant
$N\times N$ matrix $A$, in the limit $N\to\infty$. Now for fixed
$t_k$, suppose we considered the (very similar) transform:
\beq
t_k = {1\over\nu k}\tr \tA^{-k}
\eeq
that differs only by a change of sign. The point is that this
apparently harmless reversal of the $t_k$ brings about a significant
change in the matrix $A$. Moreover this is possible only in the
infinite $N$ limit, since we are trying to satisfy:
\beq
\tr A^{-k} = -\tr \tA^{-k}
\eeq
for all $k$. Now it is easy to see that the matrices $A$ and $\tA$
satisfy the following identity:
\beq
\det(\Id\otimes\Id - A^{-1}\otimes Z)=
{1\over \det(\Id\otimes\Id - \tA^{-1}\otimes Z)}
\eeq
Therefore a correlator of inverse loop operators can be rewritten in
terms of usual loop operators using:
\beq
\label{dualdet}
\left\langle{1\over \det(A\otimes \Id - \Id\otimes Z)}\right\rangle=
{1\over (\det A\tA)^N} 
\blangle\det(\tA\otimes \Id - \Id\otimes Z)\brangle
\eeq
In the light of our previous observation that inverse determinant
operators might indicate the bosonic nature of FZZT-ZZ strings at
$c=1$, it is tempting to think of \eref{dualdet} as a statement of
fermi-bose equivalence!

In terms of the operator $\det(\tA\otimes \Id - \Id\otimes Z)$,
we can make the statement that its insertion into the partially
unperturbed NMM gives rise to the KP model depending on the ``dual''
Kontsevich matrix $A$.

\section{Normal matrix model at finite $N$}

The correspondence between NMM and KP model demonstrated in
\sref{Equivalence} is valid for any $N$, as long as the parameters of
the former are restricted to a subspace. The NMM itself is supposed to
work at $N\to\infty$, in which case this restriction goes
away. However, as noted in Ref.\cite{Alexandrov:2003qk}, there is
another way to implement the NMM: by setting $N=\nu R$ (which amounts
to $N=\nu$ for $R = 1$), which they labelled as ``Model II''. In other
words, these authors argue that:
\beq
\lim\limits_{N \to \infty} Z_{NMM}(N,t,\nu) = Z_{NMM}(N=\nu R,t,\nu)
\eeq
Thus the NMM describes the $c=1$ theory at this finite value of $N$,
after analytically continuing the cosmological constant $\mu=i\nu$ to
an imaginary value\footnote{Whereas
the authors of Ref.\cite{Alexandrov:2003qk} presented this as the
analytic continuation of $N$ to the imaginary value $-i\mu$, we prefer
to think of it as continuing the cosmological constant $\mu$ to the
imaginary value $iN$.}. 

The key property of this choice is that the logarithmic term in the
matrix potential of the NMM gets tuned away.  Let us take $R=1$ from
now on. Suppose we evaluate the expectation value of the inverse
determinant operator at this $N$ (for the moment we assume that this
special value is integral). For $N$ insertions of the inverse
determinant, it gives the KP model with $N=\nu$. Thus, as one would
expect, the $\log$ term of the KP model is also tuned away. Now if we
choose $\tbar_k=c\,\dl_{k3}$, with $c$ some constant, then the KP
model reduces to the Kontsevich model, as observed in
Ref.\cite{Imbimbo:1995yv}. This shows that the Kontsevich model is a
particular deformation of the $c=1$ string theory after analytic
continuation to imaginary cosmological constant and condensation of a
particular tachyon ($T_3$). Note that at the end of this procedure,
the rank of the Kontsevich matrix is the same as that of the NMM
matrix.

As mentioned earlier, there is a different route to the Kontsevich
model starting from the Gaussian Matrix
Model (GMM)\cite{Maldacena:2004sn}. Here one starts with a Gaussian matrix
model of rank $\hat N$, with $N$ insertions of the determinant
operator, and takes ${\hat N}\to\infty$ as a double-scaling limit by
focussing on the edge of the eigenvalue distribution. The result is
the Kontsevich matrix model. This time the rank ${\hat N}$ of the
original matrix has disappeared from the picture (it was sent to
infinity) while the Kontsevich matrix inherits its rank from the
number of determinant insertions $N$.

A diagram of the situation is given in Fig.1.
\FIGURE{
\epsfxsize=11cm
\epsfbox{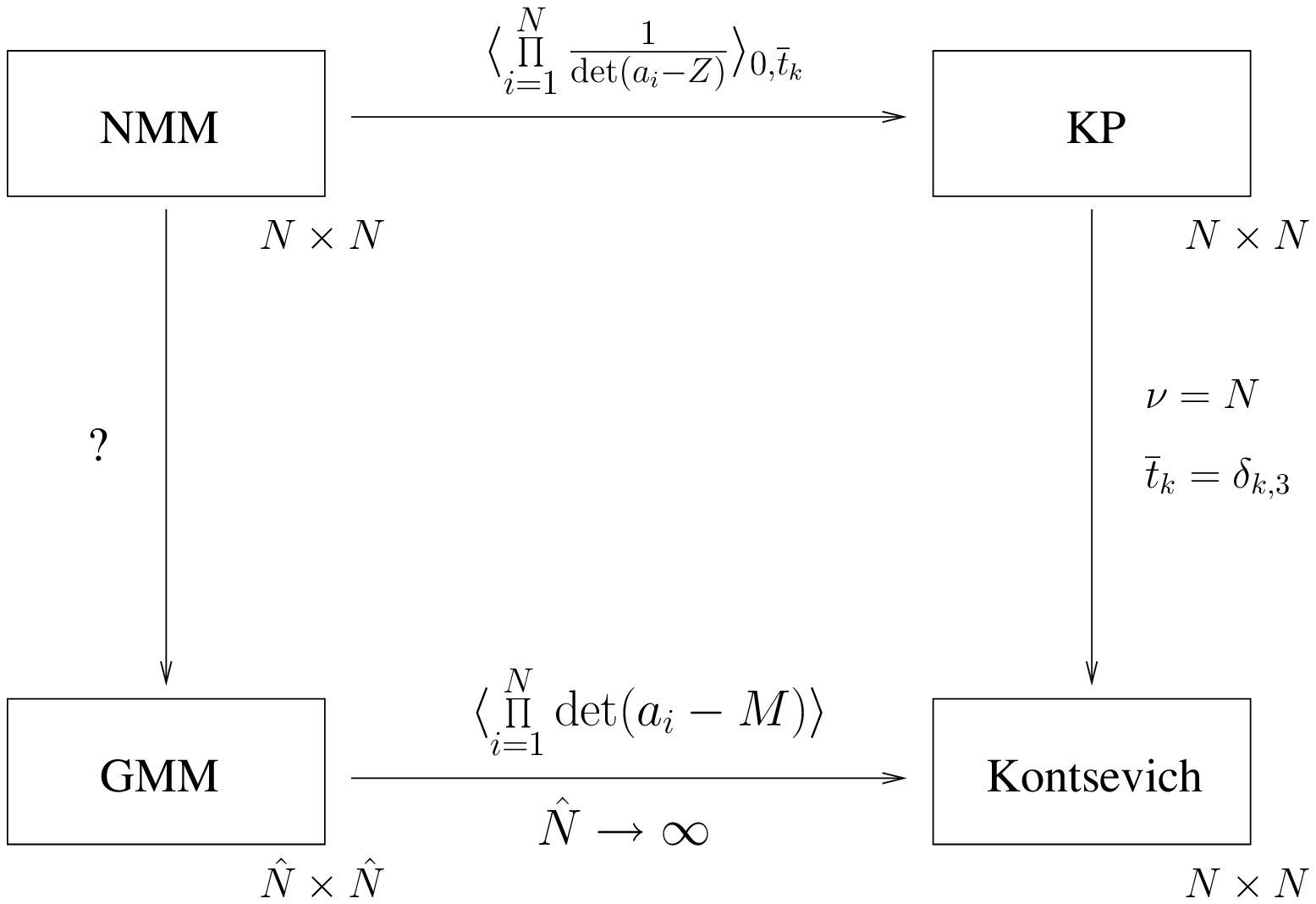}
\caption{Two routes from NMM to Kontsevich}}
>From the figure one sees that the diagram can be closed if we find a
suitable relation of the NMM to the Gaussian matrix model. This is not
hard to find at a qualitative level. In fact with $\nu=N$ and
$t_k,\tbar_k=0$ the NMM {\it is} a Gaussian matrix model. We choose
the rank to be ${\hat N}$. The NMM eigenvalue distribution
$\rho(z,\zbar)$ is constant inside a disc in the $z$-plane (for
$R=1$)\cite{Alexandrov:2003qk}. If we look at a contour along the real
axis in the $z$-plane, then the effective eigenvalue distribution
\beq
\rho(x) = \int dy~ \rho(x,y)
\eeq
is a semi-circle law, and we find the Gaussian matrix model. However,
this picture of eigenvalue distributions is valid only at large ${\hat
N}$. Inserting $N$ determinant operators and taking ${\hat
N}\to\infty$ as a double-scaling limit, one recovers the Kontsevich
model. In this way of proceeding, the cubic coupling of the Kontsevich
model is switched on automatically during the double-scaling limit. In
the alternative route through the KP model, one has to switch on the
coupling $\tbar_3$ by hand. A more detailed understanding of these two
routes and their relationship should illuminate the question of how
minimal model strings are embedded in $c=1$. We leave this for future
work.

\section{Conclusions}
\label{Conclusions}

We have established the equivalence between two matrix models of the
$c=1$ string (at selfdual radius): the Normal Matrix Model of
Ref.\cite{Alexandrov:2003qk} and the Kontsevich-Penner model of
Ref.\cite{Imbimbo:1995yv}. Both matrix models were initially found as
solutions of a Toda hierarchy, so this equivalence is not very
surprising. However, it is still helpful to have an explicit
derivation, which also uncovered a few subtleties. Also we ended up
showing why the KP matrix model does not exist at radius $R\ne 1$.

The more interesting aspect of this equivalence is that correlation
functions of inverse determinant operators in the partially
unperturbed NMM give rise to the KP model. This is analogous to
corresponding results in
Refs.\cite{Maldacena:2004sn,Hashimoto:2005bf}, with two important
differences. In those cases, one considered determinants rather than
inverse determinants, and their correlators were computed in a
double-scaled matrix model. In the NMM there is no double-scaling as
it already describes the grand canonical partition function of the
double-scaled Matrix Quantum Mechanics. Another difference is that the
$N$ of the final (KP) model is equal to that of the NMM, and part of
the matrix variables in NMM survive as the matrices of the KP
model. All this suggests that, if one makes an analogy with the
topological minimal models, the NMM occupies a position half-way
between the original matrix model arising from dynamical triangulation
of random surfaces (which requires a double-scaling limit to describe
continuum surfaces) and the final ``topological'' model. If this is
true, we may have only described half the story of open-closed duality
at $c=1$ while the correspondence between MQM and NMM constitutes the
previous half. Further work may lead to a more coherent picture of the
steps involved and thereby a deeper understanding of open-closed
string duality at $c=1$.

As we commented earlier, the inverse determinant operators seem to
suggest bosonic statistics for FZZT-ZZ branes (at least in the NMM
context) in contrast to fermionic statistics for $c<1$. Another way to
think of this is that both determinant and inverse determinant
operators expand out to give the same set of macroscopic loops, the
only difference being a minus sign for an odd number of loops in the
latter case. Alternatively one can think of the basic loop operator as
being changed by a sign to $-\tr\log(a-Z)$. Either way, the role of inverse
determinant operators clearly calls for further investigation.

We commented earlier that trying to take the $c=1$ limit of $c<1$ FZZT
correlators is problematic and therefore a derivation of the KP model
from open-string field theory analogous to Ref.\cite{Gaiotto:2003yb}
has not been forthcoming. While this may yet be achieved, the
situation recalls a historical parallel. In the 1990's, attempts to
derive $c=1$ closed string theory as a limit of the $c<1$ theories
were not very successful. Eventually it was found that at least at
selfdual radius, the $c=1$ string is a nonstandard case -- rather than
a limit -- of the $c<1$ models. This was understood by going over to
the
topological\cite{Gato-Rivera:1992ki,Bershadsky:1992qg,Mukhi:1993zb}
rather than conventional, formulation of these string theories. It
emerged that while the $(p,q)$ minimal models for varying $q$ were
described by topological models labelled by an integer $k=p-2\ge 0$
(for example, $SU(2)_k/U(1)$ twisted Kazama-Suzuki models or twisted
${\cal N}=2$ Landau-Ginzburg theories with superpotential $X^{k+2}$),
the $c=1$ string at selfdual radius was instead described by
``continuations'' of these models to
$k=-3$\cite{Witten:1991mk,Mukhi:1993zb,Ghoshal:1993qt,Hanany:1994fi},
rather than the more naive guess one might have made, namely
$k\to\infty$. Therefore progress on FZZT branes at $c=1$ in the
continuum formulation might most naturally emerge in the context of
topological D-branes in the twisted $SU(2)_{-3}/U(1)$ Kazama-Suzuki
model or $X^{-1}$ Landau-Ginzburg theory. Indeed,
Ref.\cite{Aganagic:2003qj} represents important progress in this
direction, and the Kontsevich model has been obtained there in the
topological setup, predating the more recent derivations of
Refs.\cite{Gaiotto:2003yb,Maldacena:2004sn}. In fact, the KP model of
Ref.\cite{Imbimbo:1995yv} was also obtained in
Ref.\cite{Aganagic:2003qj}. 

Extension of the NMM/KP models to include winding modes of the $c=1$
string, as well as a better understanding of 2d black holes from
matrix models\cite{Kazakov:2000pm,Alexandrov:2001cm,%
Alexandrov:2002fh,Alexandrov:2003qk}, remain open problems and perhaps
the open-closed duality studied here will be helpful in this regard.

We have not pursued here an observation made in
Ref.\cite{Mukhi:2003sz} that the KP model simplifies when we
exponentiate the matrix variable via $M=e^\Phi$. The resulting model,
which was named the ``Liouville matrix model'' there, is suggestive of
$N$ D-instantons moving in a Liouville plus linear potential. A
similar exponentiation can be carried out in the NMM. In either case
this is an almost trivial change of variables, therefore it does not
seem important for the considerations in the present paper. However,
in the light of the present work, these changes of variables might
lead to new and more satisfying interpretations of the matrix models
themselves.

As a final comment, we note that open-closed duality has in recent
times been given a more fundamental basis in the Gopakumar
programme\cite{Gopakumar:2003ns,Gopakumar:2004qb,Gopakumar:2005fx}
where closed string theory is proposed to be derived from quite
general large-$N$ field theories. Now this programme is expected to
apply not just to noncritical strings but to all string theories. We
know that the Kontsevich and Penner models compute topological
invariants of the moduli space of Riemann surfaces, but the above
works seem to suggest that these models play a role in more
complicated string theories too. If so, equivalences and open-closed
dualities such as we have discussed here may have more far-reaching
implications than just providing examples in simplified string
backgrounds.

\section*{Acknowledgements}

We are grateful to Debashis Ghoshal, Kevin Goldstein and Rajesh
Gopakumar for discussions, and the people of India for generously
supporting our research. The research of AM was supported in part by CSIR
Award No. 9/9/256(SPM-5)/2K2/EMR-I.

\end{document}